\documentclass[twocolumn]{aastex6}

\usepackage{graphicx}
\usepackage{amsmath}
\usepackage{amssymb}

\newcommand{\mbf}[1]{\mbox{\boldmath $#1$}}
\newcommand{\mbfs}[1]{\mbox{\scriptsize\boldmath $#1$}}

\newcommand{\Eqn}[1]{Equation~(\ref{eqn:#1})}

\newcommand{\eqn}[1]{equation~(\ref{eqn:#1})}
\newcommand{\eqns}[1]{equations~(\ref{eqn:#1})}

\newcommand{\Ci}{\ensuremath{i}}

\newcommand{\sech}{\ensuremath{\rm sech}\, }
\newcommand{\var}{{\rm var}}
\newcommand{\trace}{{\rm tr}}
\newcommand{\real}{{\rm Re}}
\newcommand{\imag}{{\rm Im}}

\newcommand{\Rotation}{{\bf R}}
\newcommand{\Boost}{{\bf B}}

\newcommand{\vRotation}[1][n]{\ensuremath{\Rotation_{\mbfs{\hat #1}}}}
\newcommand{\vBoost}[1][m]{\ensuremath{\Boost_{\mbfs{\hat #1}}}}

\newcommand{\rotat}{\ensuremath{\vRotation(\phi)}}
\newcommand{\boost}{\ensuremath{\vBoost(\beta)}}

\newcommand{\pauli}[1]{\ensuremath{\mbf{\sigma}_{#1}}}

\newcommand{\Rmult}{\ensuremath{R_{\varphi{\bf J}}}}
\newcommand{\conderr}{\ensuremath{\hat\sigma_{\varphi|{\bf J}}}}

\newcommand{\psr}{PSR\,J0437$-$4715}

\shorttitle   {Matrix Template Matching}
\shortauthors {W. van Straten}

\begin{document}

% REF 1
\title{Radio Astronomical Polarimetry and High-Precision Pulsar Timing}

\author{W. van Straten}

\affil{Center for Gravitational Wave Astronomy, 
The University of Texas at Brownsville, \\
80 Fort Brown, Brownsville, TX 78520}

\email{vanstraten.willem@gmail.com}

\begin{abstract}

A new method of matrix template matching is presented in the context
of pulsar timing analysis.  Pulse arrival times are typically measured
using only the observed total intensity light curve.  The new
technique exploits the additional timing information available in the
polarization of the pulsar signal by modeling the transformation
between two polarized light curves in the Fourier domain.  For a
number of millisecond pulsars, arrival time estimates derived from
polarimetric data are predicted to exhibit greater precision and
accuracy than those derived from the total intensity alone.
Furthermore, the transformation matrix produced during template
matching may be used to calibrate observations of other point sources.
Unpublished supplementary material is appended after the bibliography.

\end{abstract}


\section {Introduction}

High-precision pulsar timing is a well-established technique of modern
astrophysics; it has yielded the strongest constraints on theories of
gravitation in the strong-field regime~\citep{sta04a}, and is
anticipated to provide a direct detection of the stochastic
gravitational wave background due to supermassive black hole
binaries~\citep{jhlm05}.  Fundamental to every pulsar timing
experiment is a measurement known as the pulse time-of-arrival (TOA),
the epoch at which a fiducial phase of the pulsar's periodic signal is
received at the observatory.  The confidence limits of the various
physical parameters that are derived from the TOA data depend
upon the precision and accuracy with which arrival times can be
estimated.

In addition to typical constraints such as the system temperature,
instrumental bandwidth, and integration length, TOA precision also
depends upon the physical properties of the observed pulsar, including
its total flux density, pulse period, and the shape of its
phase-resolved light curve, or pulse profile.  When fully resolved,
narrow features in the pulse profile provide strong constraints during
template matching, thereby yielding greater arrival time precision
% REF 5a
(see \S~\ref{sec:precision}).  The polarized component of the pulsar
signal often displays much sharper features than observed in the total
intensity, especially when the pulsar exhibits transitions between
orthogonally polarized modes.  This important property may be
exploited to significantly improve timing precision by incorporating
polarization data.

Pulsar polarization also impacts on timing accuracy \citep{ckl+04}.
The total intensity profile can be significantly distorted by
instrumental artifacts, a problem most readily observed as a
systematic variation of arrival time residuals with parallactic
angle~\citep{van03}.  To address this issue, \cite{bri00} proposed
timing the polarimetric invariant profile, which greatly improved the
timing accuracy of \psr~\citep{bvb+00}.  However, there are
disadvantages of using the invariant profile; its signal-to-noise
% REF 6c
ratio (S/N) is at most $1/\sqrt{2}$ times that of the total intensity
(see \S~\ref{sec:application}) and inversely proportional to the
degree of polarization; also, its computation suffers from imprecision
in the estimation of the off-pulse signal.  Therefore, use of the
invariant profile can be detrimental when the pulsar is highly
polarized or has low flux density.

In contrast, the matrix template matching technique presented in this
paper can be used to improve both the precision and accuracy of
arrival time estimates without any sensitivity to mean off-pulse
polarization.  The method simultaneously yields the polarimetric
transformation between the template and observed profiles, which may
be utilized to completely calibrate the instrumental response in other
observations.
Following a review of the required mathematics in
\S~\ref{sec:jones}, the matrix template matching technique is
formulated and quantitatively compared with scalar methods in
\S~\ref{sec:modeling}.  In \S~\ref{sec:application}, the analysis is
demonstrated using a sample of millisecond pulsars.  The application
of matrix template matching to instrumental calibration is outlined in
\S~\ref{sec:calibration}, and the main conclusions of this work are
summarized in \S~\ref{sec:conclusions}.

\section{Review of the Jones Calculus}
\label{sec:jones}

% REF 7a
The formulation and analysis of the matrix template matching method
utilizes the terminology and notation reviewed in this section.
% REF 7b
The polarization of electromagnetic radiation is described by the
second-order statistics of the transverse electric field vector,
$\mbf{e}$, as represented using the complex $2\times2$ coherency
matrix, $\mbf{\rho}=\langle\mbf{e \otimes e}^\dagger\rangle$
% REF 7c
\citep{bw80}.  Here, the angular brackets denote an ensemble average,
$\mbf\otimes$ is the matrix direct product, and $\mbf{e}^\dagger$ is
the Hermitian transpose of $\mbf{e}$.  The coherency matrix is
commonly expressed as a linear combination of Hermitian basis
matrices, ${\mbf\rho}=(S_0\,\pauli{0}+\mbf{S\cdot\sigma})/2$, where
$\pauli{0}$ is the $2\times2$ identity matrix, $\mbf{\sigma} =
(\pauli{1},\pauli{2},\pauli{3})$ are the Pauli spin matrices, $S_0$ is
the total intensity, and $\mbf{S} = (S_1,S_2,S_3)$ is the polarization
vector \citep{bri00}.  The Pauli matrices are traceless and satisfy
$\pauli{i}^2=\pauli{0}$; therefore, $S_k =
\trace(\pauli{k}\mbf{\rho})$, where $\trace$ is the matrix trace
operator \citep{ham00}.

% REF 7d
In the analysis of the reception of polarized radiation, the following
conventions are used.  The response of a single receptor is defined by
the Jones vector, $\mbf{r}$, such that the voltage induced in the
receptor by the incident electric field is given by the scalar
product, $v=\mbf{r}^\dagger\mbf{e}$.  A dual-receptor feed is
represented by the Hermitian transpose of a Jones matrix with columns
equal to the Jones vector of each receptor,
\begin{equation}
\label{eqn:feed}
{\bf J}=(\mbf{r}_0\; \mbf{r}_1)^\dagger =
  \left( \begin{array}{cc}
    r_{00}^* & r_{01}^* \\
    r_{10}^* & r_{11}^*
  \end{array} \right).
\end{equation}
The receptors in an ideal feed respond to orthogonal senses of
polarization (ie. the scalar product, $\mbf{r}_0^\dagger\mbf{r}_1=0$)
and have identical gains
(ie. $\mbf{r}_0^\dagger\mbf{r}_0=\mbf{r}_1^\dagger\mbf{r}_1$).  

A more meaningful geometric interpretation of Jones vectors is provided by
the corresponding Stokes parameters,
$S_k=\trace[\pauli{k}\mbf{r \otimes r}^\dagger]$.
The state of polarization to which a receptor maximally responds is
completely described by the three components of its associated Stokes
polarization vector, $\mbf{S}$.  Therefore, it is common to define a
receptor using the spherical coordinates of $\mbf{S}$ \citep{cha60};
in the linear basis, these include the gain,
$g=|\mbf{r}|=|\mbf{S}|^{1\over2}$, the orientation,
\begin{equation}
\label{eqn:orientation}
\theta = {1\over2} \tan^{-1} {S_2 \over S_1},
\end{equation}
and the ellipticity,
\begin{equation}
\label{eqn:ellipticity}
\epsilon = {1\over2} \sin^{-1} {S_3 \over |\mbf{S}|},
\end{equation}
such that
\begin{equation}
\label{receptor}
\mbf{r}=g\left( \begin{array}{c}
    \cos\theta\cos\epsilon + \Ci\sin\theta\sin\epsilon \\
    \sin\theta\cos\epsilon - \Ci\cos\theta\sin\epsilon
  \end{array} \right).
\end{equation}

% REF 7e
The impact of non-ideal feed receptors on pulsar timing is analyzed
by exploiting a powerful classification of Jones matrices motivated 
by the polar decomposition.
Any non-singular matrix can be decomposed into the product of a
unitary matrix and a positive-definite Hermitian matrix.  Using the
axis-angle parameterization \citep{bri00}, the polar decomposition of
a Jones matrix \citep{ham00} is expressed as
\begin{equation}
\label{eqn:polar}
{\bf J} = J \, \boost \, \rotat,
\end{equation}
where $J=(\det{\bf J})^{1\over2}$, \boost\ is positive-definite Hermitian,
and \rotat\ is unitary; both \boost\ and \rotat\ are unimodular.  The
unit 3-vectors, $\mbf{\hat m}$ and $\mbf{\hat n}$, correspond to axes
of symmetry in the three-dimensional space of the Stokes polarization
vector.  Under the congruence transformation of the coherency matrix,
the Hermitian matrices,
\begin{equation}
\label{eqn:Boost}
\boost = \exp (\beta\,\mbf{\hat{m}\cdot\sigma})
       = \pauli{0}\cosh\beta + \mbf{\hat{m}\cdot\sigma}\sinh\beta,
\end{equation}
effect a Lorentz boost of the Stokes 4-vector along the
$\mbf{\hat m}$ axis by an impact parameter $2\beta$, such that the
resulting total intensity, $S^\prime_0 = S_0 \cosh 2\beta +
\mbf{S\cdot\hat{m}} \sinh 2\beta$.  Similarly, the unitary matrices,
\begin{equation}
\label{eqn:Rotation}
\rotat = \exp (\Ci\phi\,\mbf{\hat{n}\cdot\sigma})
       = \pauli{0}\cos\phi + \Ci\mbf{\hat{n}\cdot\sigma}\sin\phi,
\end{equation}
rotate the Stokes polarization vector about the $\mbf{\hat n}$ axis by
an angle $2\phi$, leaving the total intensity unchanged.  This
parameterization enables the important distinction between
polarimetric transformations that mix the total and polarized
intensities (boosts) and those that effect a change of basis
(rotations).

\section{Matrix Template Matching}
\label{sec:modeling}

% REF 8b
The primary purpose of this paper is the formal description of the
matrix template matching technique and the quantitative comparison of
its effectiveness with that of conventional scalar methods.  The
performance of a method of TOA measurement may be evaluated in terms
of the precision and accuracy of the arrival time estimates that it
produces.  The analysis of TOA precision requires careful attention to
the propagation of experimental error, as described in
\S~\ref{sec:precision}.  TOA accuracy depends upon the susceptibility
of the technique to sources of systematic error.  To quantitatively
compare different methods, a simulation that spans the full range of
potential artifacts is devised in \S~\ref{sec:accuracy}.

\subsection{Description of Technique}
\label{sec:description}

A pulsar's mean pulse profile is measured by integrating the observed
flux density as a function of pulse phase.  By averaging many pulse
profiles, one with high S/N may be formed and used as a template
against which the individual observations are matched.  The best-fit
phase shift derived by the template matching procedure is then used to
compute the pulse TOA.

\cite{tay92} presents a method for modeling the phase shift between
the template and observed total intensity profiles in the Fourier
domain.  In the current treatment, the scalar equation that relates
two total intensity profiles is replaced by an analogous matrix
equation, which is expressed using the Jones calculus.  Let the
coherency matrices, $\mbf{\rho}^\prime (\phi_n)$, represent the
observed polarization as a function of discrete pulse phase, $\phi_n$,
where $0\le n< N$ and $N$ is the number of intervals into which the
pulse period is evenly divided.  Each observed polarization profile is
related to the template, $\mbf{\rho}_0(\phi_n)$, by the matrix
equation,
\begin{equation}
\label{eqn:model}
\mbf{\rho}^\prime (\phi_n) = 
        {\bf J} \mbf{\rho}_0 (\phi_n - \varphi) {\bf J}^\dagger
	+ \mbf{\rho}_{\mathrm DC} + \mbf{\rho}_{\mathrm N}(\phi_n),
\end{equation}
where ${\bf J}$ is the polarimetric transformation, $\varphi$ is the
phase shift, $\mbf{\rho}_{\mathrm DC}$ is the DC offset between the
two profiles, and $\mbf{\rho}_{\mathrm N}$ represents the system
noise. The discrete Fourier transform (DFT) of \eqn{model} is
\begin{equation}
\mbf{\rho}^\prime (\nu_m) = 
  {\bf J} \mbf{\rho}_0 (\nu_m) {\bf J}^\dagger \exp(-i2\pi \nu_m \varphi)
  + \mbf{\rho}_{\mathrm N}(\nu_m),
\label{eqn:fourier_rho}
\end{equation}
where $\nu_m$ is the discrete frequency.  Given the observed
Stokes parameters, $S_k^\prime(\phi_n)$, and their DFTs,
$S_k^\prime(\nu_m)$, the best-fit model parameters will minimize
the objective merit function,
\begin{equation}
\chi^2 = \sum_{m=1}^{N/2} \sum_{k=0}^3 { |S_k^\prime(\nu_m) -
\trace[\pauli{k}\;\mbf{\rho}^\prime(\nu_m)]|^2 \varsigma_k^{-2} },
\label{eqn:merit}
\end{equation}
where $\varsigma_k$ is equal to the rms of the noise in each DFT and
$\trace$ is the matrix trace operator.  As in~\cite{van04}, the
partial derivatives of \eqn{merit} are computed with respect to both
$\varphi$ and the seven non-degenerate parameters that determine ${\bf
J}$.  The Levenberg-Marquardt method is then applied to find the
parameters that minimize $\chi^2$ \citep{ptvf92}.

\subsection{Timing Precision}
\label{sec:precision}

When compared with the scalar technique, matrix template matching
quadruples the number of observational constraints while introducing
only six degrees of freedom.  Therefore, arrival time estimates
derived from the polarization profile might be expected to have
greater precision than those derived from the total intensity profile
alone.  However, the effectiveness of matrix template matching depends
upon both the degree of polarization and the variability of the
polarization vector as a function of pulse phase.  These properties
determine the extent to which the phase shift, $\varphi$, is
correlated with the free parameters that determine the Jones matrix,
{\bf J}.

To estimate the timing precision attainable by matrix template
matching, consider the solution of \eqn{model} in the special case
that {\bf J} is known.  After calibration, the minimization of
\eqn{merit} by variation of $\varphi$ requires finding the appropriate
root of
\begin{equation}
\label{eqn:delchisq_delvarphi}
{\partial\chi^2\over\partial\varphi} =
 4\pi \sum_{m=1}^{N/2} \sum_{k=0}^3 { |S_{k,m}| \over \varsigma_k^2 } 
 \nu_m \sin ( \phi_{k,m} - 2\pi\varphi\nu_m ),
\end{equation}
where
$S_{k,m}=S_k^{\prime*}(\nu_m)\trace[\pauli{k}\;\mbf{\rho}_0(\nu_m)]$
is the cross-spectral power of the template and observation and
$\phi_{k,m}$ is the argument of $S_{k,m}$.  To first order, 
\begin{equation}
\label{eqn:first_order}
{\partial\chi^2\over\partial\varphi} \simeq
 4\pi \sum_{m=1}^{N/2} \sum_{k=0}^3 { |S_{k,m}| \over \varsigma_k^2 } 
 (\phi_{k,m}\nu_m  - 2\pi\varphi\nu_m^2) = 0,
\end{equation}
which is equivalent to the solution of a line with slope $2\pi\varphi$
passing through the origin and the points, $(\nu_m, \phi_{k,m})$.
This linear approximation is readily solved for $\varphi$ and used to
determine the conditional variance,
\begin{equation}
\label{eqn:conditional_variance}
\var(\varphi|{\bf J})
  = \left[ 4\pi^2 \sum_{m=1}^{N/2} \nu_m^2 
           \sum_{k=0}^3 {|S_{k,m}| \over \varsigma_k^2 } \right]^{-1}.
\end{equation}
Equation~\ref{eqn:conditional_variance} demonstrates that fluctuation
power contributes quadratically as a function of frequency to the
reduction of the phase shift variance.  
% REF 5b
In other words, sharper features in the pulse profile, which generate
more power at higher harmonics, yield greater arrival time precision.
This general results holds for both matrix and scalar template
matching (in the scalar case, the sum over index $k$ stops at 0).

Equation~\ref{eqn:conditional_variance} also provides an upper limit
on the arrival time precision that may be obtained in the ideal case
of a perfectly calibrated instrument.  However, the instrumental 
response is generally unknown, and the Jones matrix must also be
varied in order to minimize \eqn{merit}.
% REF 8c
The covariances between the phase shift, $\varphi$, and the free
parameters that describe {\bf J} will increase the uncertainty in
$\varphi$ and therefore decrease arrival time precision.

Formally, the variance of $\varphi$ is determined by the covariance
matrix of the free parameters, ${\bf C}={\mbf\alpha}^{-1}$, where
$\mbf\alpha$ is the curvature matrix.  Assuming that the noise power
in the DFT of each Stokes parameter is equal,
\begin{equation}
\label{eqn:curvature}
\alpha_{rs} = {1\over2}{\partial^2\chi^2\over\partial\eta_r\partial \eta_s} 
= {2 \over \varsigma^2} \sum_{m=1}^{N/2}
  \real\left[
    \trace\left( {\partial\mbf{\rho}^{\prime\dagger}_m\over\partial\eta_r}
                 {\partial\mbf{\rho}^\prime_m\over\partial\eta_s} 
          \right)
       \right].
\end{equation}
If the free parameters, $\mbf{\eta}$, are partitioned into
$\varphi=\eta_0$ and the seven Jones matrix parameters,
$\mbf{\eta}_{\bf J}$, then {\bf C} may be conformably partitioned into
\citep{soa99}
\begin{equation}
\label{eqn:partition}
{\bf C} = \left( \begin{array}{cc}
c_{\varphi\varphi} &  {\mbf C}_{\varphi{\bf J}}^T \\
{\mbf C}_{\varphi{\bf J}} & {\bf C_{JJ}}
\end{array}\right),
\end{equation}
and $\var(\varphi)=c_{\varphi\varphi}$.  Furthermore, if the free
parameters are multinormally distributed, then the multiple
correlation between $\varphi$ and $\mbf{\eta}_{\bf J}$,
\begin{equation}
\label{eqn:multiple_correlation}
\Rmult = \left[c_{\varphi\varphi}^{-1}{\mbf C}_{\varphi{\bf J}}^T
               {\bf C_{JJ}}^{-1}{\mbf C}_{\varphi{\bf J}}\right]^{1\over2},
\end{equation}
describes the relationship between the variance and conditional
variance of $\varphi$,
\begin{equation}
\label{eqn:variance}
\var(\varphi|{\bf J})=\var(\varphi)(1-R^2_{\varphi{\bf J}}).
\end{equation}
The multiple correlation coefficient, $0\le\Rmult\le1$, provides a
useful measure of the decrease in arrival time precision that results
from an unknown instrumental response.  
% REF 3a
It is important to note that \Rmult\ can be computed using only the
template polarization profile; that is, though $\mbf\alpha$ depends on
$\mbf{\rho}^\prime_m$ and $\varsigma$, all factors of $\varsigma$ are
canceled in \eqn{multiple_correlation}.  Therefore, \Rmult\ is unique
to each pulsar and is independent of the S/N of the observations.

% REF 3b
However, the precision and accuracy with which \Rmult\ can be
estimated does depend on the S/N of the template.  In fact,
measurement noise in the template polarization profile artificially
decreases both the multiple correlation and the conditional variance
of $\varphi$.  To minimize the impact of noise on the computation of
$\var(\varphi|{\bf J})$, \Rmult, and $\var(\varphi)$, the summations
over the index $m$ in \eqns{conditional_variance}
and~(\ref{eqn:curvature}) are performed up to a maximum harmonic, $M$,
the highest frequency at which the fluctuation power spectra exhibit
three consecutive harmonics with power greater than three times the
mean noise power.

To completely eliminate the dependence on $\varsigma$,
$\var(\varphi|{\bf J})$ and $\var(\varphi)$ are normalized by the
corresponding variance in the phase shift, $\bar\varphi$, yielded by
the conventional method of scalar template matching the total
intensity profile.  That is, \eqn{conditional_variance} is used to
predict the relative conditional timing error in the case of a known
instrumental response,
\begin{equation}
\label{eqn:conderr}
\conderr
= \left[{\var(\varphi|{\bf J})\over\var(\bar\varphi|G)}\right]^{1\over2}
= \left[{\Sigma_0\over\Sigma_0+\Sigma_{1-3}}\right]^{1\over2},
\end{equation}
where $G$ is the absolute gain,
\begin{equation}
\Sigma_0 = \sum_{m=1}^M\nu_m^2|S_0(\nu_m)|^2
\end{equation}
and
\begin{equation}
\Sigma_{1-3} = \sum_{m=1}^M\nu_m^2|\mbf{S}(\nu_m)|^2.
\end{equation}
Similarly,
\begin{equation}
\label{eqn:relerr}
\hat\sigma_{\varphi} = \left({c_{\varphi\varphi} \over
                              c_{\bar\varphi\bar\varphi}} \right)^{1\over2}
\end{equation}
yields the relative timing error in the general case of an unknown
response. Given a template polarization profile,
\eqns{conditional_variance} -- (\ref{eqn:relerr}) can be used to predict
arrival time uncertainty in the case of either known or unknown
polarimetric response.
% REF 8d
These equations provide the basis for comparing the precision of
scalar and matrix template matching methods for each pulsar in
\S~\ref{sec:application}.

% REF 3c
Note that $\conderr \leq 1$; that is, for a known response, the
precision yielded by matrix template matching will always be as good
as or better than that of the conventional scalar method.  In the
general case of unknown response, matrix template matching will yield
more precise TOA estimates only when $\det\mbf{\alpha}\neq0$ and
\begin{equation}
\label{eqn:restrict}
\Rmult^2 + \conderr^2 < 1.
\end{equation}
The interpretation of the first restriction is simplified by
considering the polar decomposition of the unknown response, {\bf J}.
To constrain the rotation component of this matrix, the pulse profile
must include at least two non-collinear polarization vectors.
Therefore, $\mbf\alpha$ becomes singular when either the profile is
unpolarized or the polarization vectors, $\mbf{S}(\phi_n)$, lie along
a single line.  In these two special cases, it is not possible to
invert $\mbf\alpha$ and matrix template matching fails.

\Eqn{restrict} places an upper limit on the value of \Rmult.
As the degree of polarization of the pulse profile approaches zero,
\conderr\ approaches unity, and the maximum value of \Rmult\ permitted
by \eqn{restrict} approaches zero. 
Owing to the unknown boost component of the response, the multiple
correlation approaches unity when there is a high degree of symmetry in 
$S_0(\phi_n)$ combined with antisymmetry in $\mbf{S}(\phi_n)$.  
Similarly, axial symmetry in $\mbf{S}(\phi_n)$ will increase the
multiple correlation between the rotation component and $\varphi$.
In these cases, the uncertainty of arrival time estimates derived
from the polarization profile will be larger than that of those 
derived from the total intensity alone.

In summary, matrix template matching does not perform as well as the
conventional method when the degree of polarization is low and the
multiple correlation (due to the symmetry properties of the
polarization profile) is high.  However, in general, the radiation
from pulsars is highly polarized and, especially in the millisecond
pulsar population, most polarization profiles exhibit complex
structure.  Therefore, it is expected that, in all but a few
exceptional cases, matrix template matching will produce arrival time
estimates with greater precision than those derived by conventional
scalar methods.

\subsection{Timing Accuracy}
\label{sec:accuracy}

Unmodeled polarimetric distortion of the total intensity profile can
shift arrival time estimates derived by conventional methods.  As
\eqn{model} incorporates an unknown polarimetric transformation, it
has the capacity to model instrumental polarization instabilities and
isolate them from pulse arrival time variations, thereby eliminating a
potentially significant source of systematic timing error.
% REF 8e
The improvement in timing accuracy may be analyzed by designing a
simulation in which arrival time estimates are derived from
polarization profiles that have been subjected to transformations that
distort the total intensity profile.

As reviewed in \S~\ref{sec:jones}, only boost transformations affect
the total intensity.  Physically, boosts arise from the differential
amplification and non-orthogonality of the feed receptors \citep[and
references therein]{bri00}. A brief review of these phenomena provides
the motivation for the analysis technique presented in this section.

For a pair of orthogonal receptors with different gains, $g_0$ and
$g_1$, define the orthonormal receptors, $\mbf{\hat
r}_0=\mbf{r}_0/g_0$ and $\mbf{\hat r}_1=\mbf{r}_1/g_1$, and substitute
into \eqn{feed} to yield
\begin{equation}
\label{eqn:unequal_gain}
{\bf J}=(\mbf{r}_0\; \mbf{r}_1)^\dagger = G
  \left( \begin{array}{cc}
    \Gamma & 0 \\
    0   & \Gamma^{-1}
  \end{array} \right) (\mbf{\hat r}_0\; \mbf{\hat r}_1)^\dagger,
\end{equation}
where $G=(g_0g_1)^{1\over2}$ is the absolute gain and
$\Gamma=(g_0/g_1)^{1\over2}$ parameterizes the differential gain
matrix.  \Eqn{unequal_gain} is a polar decomposition and, by
substituting $\Gamma=\exp(\beta)$, the differential gain matrix may be
expressed in the form of \eqn{Boost},
\begin{equation}
\label{eqn:diff_gain}
\boost =
  \left( \begin{array}{cc}
    e^\beta & 0 \\
    0   & e^{-\beta}
  \end{array} \right) = \pauli{0}\cosh\beta + \pauli{1}\sinh\beta,
\end{equation}
where $\mbf{\hat m}=(1,0,0)$ and\footnote{There is an error following
equation~(14) in \cite{bri00}, where it should read
$\beta=\ln(g_a/g_b)/2$.}
\begin{equation}
\label{eqn:differential_gain}
\beta = {1\over2} \ln {g_0 \over g_1}.
\end{equation}
In the case of linearly polarized receptors, the $\mbf{\hat m}$ axis
lies in the Stokes $Q$-$U$ plane; for circularly polarized receptors,
$\mbf{\hat m}$ corresponds to $\pm$ Stokes $V$.  To first order,
$\beta=\gamma/2$, where $\gamma=g_0/g_1-1$ is the differential gain
ratio.

For a pair of non-orthogonal receptors, first consider the spherical 
coordinate system introduced in \S~\ref{sec:jones}.  The orientations 
and ellipticities of orthogonal receptors satisfy 
$\theta_0-\theta_1=\pm\pi/2$ and $\epsilon_0=-\epsilon_1$.
If $\delta_\theta$ and $\delta_\epsilon$
parameterize the departure from orthogonality in each of these angles,
such that $\theta_0^\prime=\theta_0+\delta_\theta$ and
$\epsilon_0^\prime=\epsilon_0+\delta_\epsilon$, then
\begin{equation}
\label{eqn:scalar}
\mbf{r}_0^{\prime\dagger}\mbf{r}_1 = g_0g_1
  (\sin\delta_\theta \cos(\delta_\epsilon+2\epsilon_0)
   + \Ci \cos\delta_\theta \sin\delta_\epsilon).
\end{equation}
However, this description is specific to linearly polarized receptors.
In the circular basis, $\epsilon_0=\pi/4$, and the non-orthogonality
is completely described by $\delta_\epsilon$; $\delta_\theta$ becomes
degenerate with the differential phase of the receptors.  As such a
degeneracy exists at the poles of any spherical coordinate system, a
basis-independent parameterization of non-orthogonality is sought.

To this end, it proves useful to consider the geometric relationship
between the Stokes polarization vector of each receptor.  In
particular, the scalar product,
\begin{equation}
\label{eqn:dot_product}
\mbf{S}_0\mbf{\cdot}\mbf{S}_1 
 = 2|\mbf{r}_0^\dagger\mbf{r}_1|^2 - |\mbf{r}_0|^2|\mbf{r}_1|^2,
\end{equation}
shows that orthogonally polarized receptors have anti-parallel Stokes
polarization vectors ($\mbf{S}_0\mbf{\cdot}\mbf{S}_1 =
-|\mbf{S}_0||\mbf{S}_1|$).  Furthermore, where $\Theta$ is the angle
between $\mbf{S}_0$ and $\mbf{S}_1$, the angle $\delta=(\pi-\Theta)/2$
parameterizes the magnitude of the receptor non-orthogonality, such that
\begin{equation}
\label{eqn:sin_delta}
\sin\delta = {|\mbf{r}_0^\dagger\mbf{r}_1|\over|\mbf{r}_0||\mbf{r}_1|}.
\end{equation}

It is much simpler to relate $\delta$ to the boost transformation that
results from non-orthogonal receptors.  To determine the boost component 
of an arbitrary matrix, ${\bf J}$, the polar decomposition
(eq.~[\ref{eqn:polar}]) is multiplied by its Hermitian transpose to
yield
\begin{equation}
\label{eqn:Boost2}
{\bf J}{\bf J}^\dagger = | \det{\bf J}\, |\, \vBoost^2(\beta) 
                       = | \det{\bf J}\, |\, \vBoost(2\beta).
\end{equation}
For a pair of receptors with gain, $g$, substitution of \eqn{feed}
into \eqn{Boost2} yields
\begin{equation}
\label{eqn:nonorthogonal_jones}
{\bf J}{\bf J}^\dagger =
  \left( \begin{array}{cc}
g^2 & W \\
W^* & g^2
  \end{array} \right),
\end{equation}
where $W=\mbf{r}_0^\dagger\mbf{r}_1$.  Substitute
$W=g^2e^{-\Ci\Phi}\tanh2\beta$, so that
$| \det{\bf J}\, | = \det({\bf J}{\bf J}^\dagger)^{1\over2} 
= (g^4-|W|^2)^{1\over2} = g^2 \sech2\beta$, and
\begin{equation}
\label{eqn:nonorthogonal_boost}
\vBoost(2\beta) = {{\bf J}{\bf J}^\dagger\over|\det{\bf J}\,|}
  = \pauli{0}\cosh2\beta + \mbf{\hat{m}\cdot\sigma}\sinh2\beta,
\end{equation}
where $\mbf{\hat m}=(0,\cos\Phi,\sin\Phi)$ and
\begin{equation}
\label{eqn:nonorthogonal_receptors}
\beta = {1\over2}\tanh^{-1}{|\mbf{r}_0^\dagger\mbf{r}_1|\over g^2}.
\end{equation}
To first order, $\beta\sim\delta/2$ (see eq.~[\ref{eqn:sin_delta}]),
which is consistent with the approximation in equation~(19)
of \cite{bri00}.

Due to the combined effects of differential gain and receptor
non-orthogonality, the boost axis, $\mbf{\hat m}$, can have an
arbitrary orientation.  For example, to first order in the polar
coordinate system best-suited to the linear basis, $\mbf{\hat
m}\propto(\gamma,\delta_\theta,\delta_\epsilon)$.  Furthermore, the
instrumental boost can vary as a function of both time and frequency
for a variety of reasons.  For example, the parallactic rotation of
the receiver feed during transit of the source changes the orientation
of $\mbf{\hat m}$ with respect to the equatorial coordinate system.
Also, to keep the signal power within operating limits, some
instruments employ active attenuators that introduce differential gain
fluctuations on short timescales.  Furthermore, the mismatched
responses of the filters used in downconversion typically lead to
variation of $\gamma$ as a function of frequency (see
\S~\ref{sec:calibration}).

The impact of these variations on conventional timing accuracy may be
estimated through a simulation in which copies of the template
polarization profile are subjected to a boost transformation before
the phase shift, $\bar\varphi$, between the template and the distorted
copy is measured (using only the total intensity).  By varying the
orientation of the boost axis, $\mbf{\hat m}$, the maximum and minimum
phase shift offsets for a given level of distortion, $\beta$, are
found and used to define the systematic timing error,
\begin{equation}
\label{eqn:distortion}
\Delta\tau(\beta) = P(\max\{\bar\varphi(\beta,\mbf{\hat m})\}
  - \min\{\bar\varphi(\beta,\mbf{\hat m})\}),
\end{equation}
where $P$ is the pulsar spin period.  Empirical observations made
during testing show that, to first order, $\Delta\tau$ varies linearly
with $\beta$.

\section{Application to Selected Pulsars}
\label{sec:application}

To demonstrate the results of the previous sections, the analysis is
applied to a sample of millisecond pulsars among the best for
high-precision timing experiments.  Polarization profiles with high
S/N were obtained for PSR~J0437$-$4715 \citep{van05}, PSR~J1022+1001,
PSR~J1713+0747, PSR~J1909$-$3744 \citep{ovhb04}, PSR~B1855+09
(I.~Stairs 2005, private communication) and PSR~B1937+21
\citep{stc99}.  For each pulsar, the fluctuation power spectra of the
total intensity, $|S_0(\nu_m)|^2$, and polarization,
$|\mbf{S}(\nu_m)|^2$, are plotted in Figure~1. The maximum harmonic
used to compute the arrival time uncertainty for each pulsar (as
described in \S~\ref{sec:precision}) is indicated by a vertical line
in each panel.  Over some frequency intervals, the polarization
fluctuation power of PSR~J0437$-$4715 and PSR~J1022$+$1001
exceeds that of the total intensity.  Therefore, it is expected that
out of the selected sample these two pulsars will benefit most from
the application of matrix template matching.

\begin{figure*}
\label{fig:snr}
\begin{tabular}{ll}
\includegraphics[angle=-90,width=86mm]{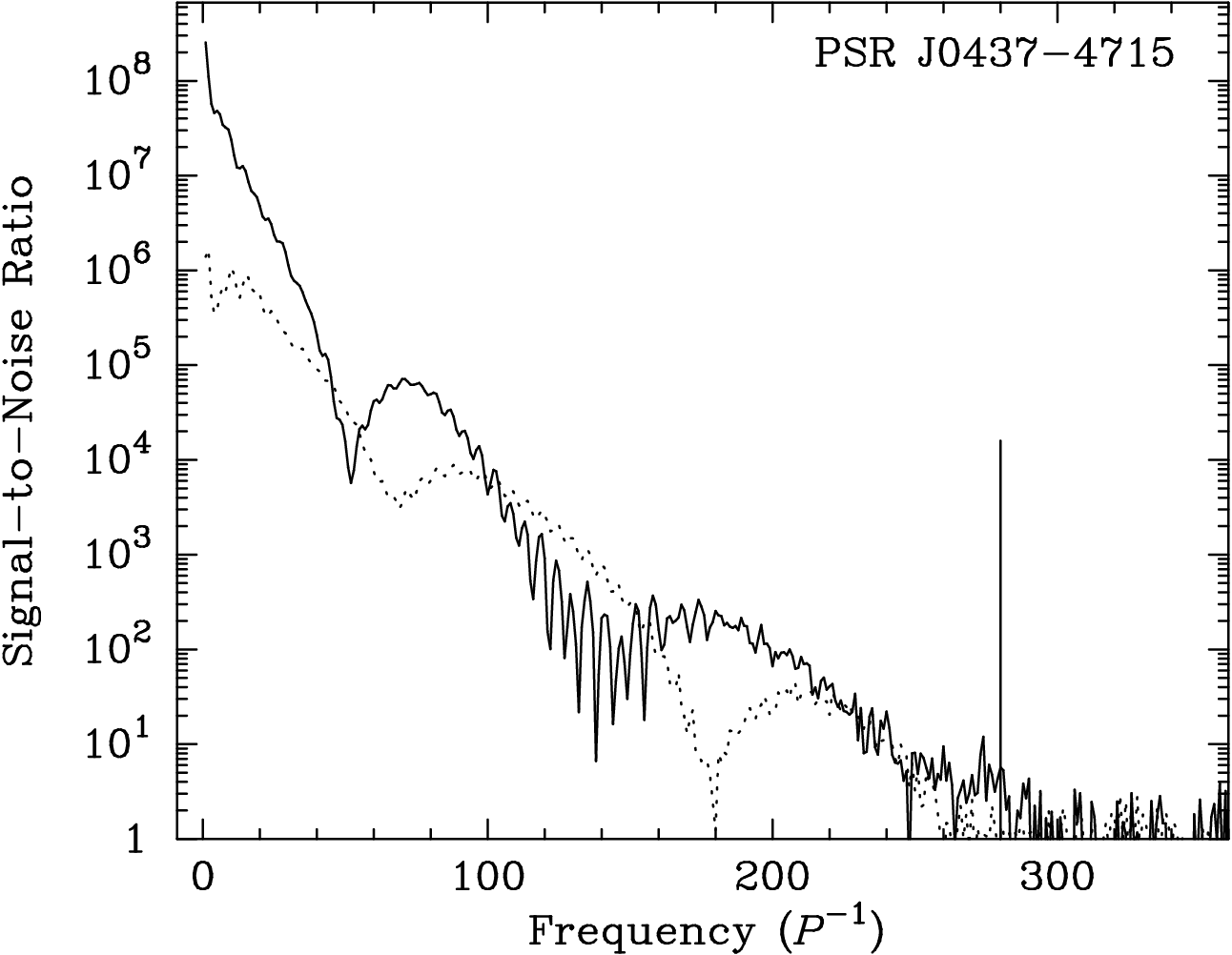} &
\includegraphics[angle=-90,width=86mm]{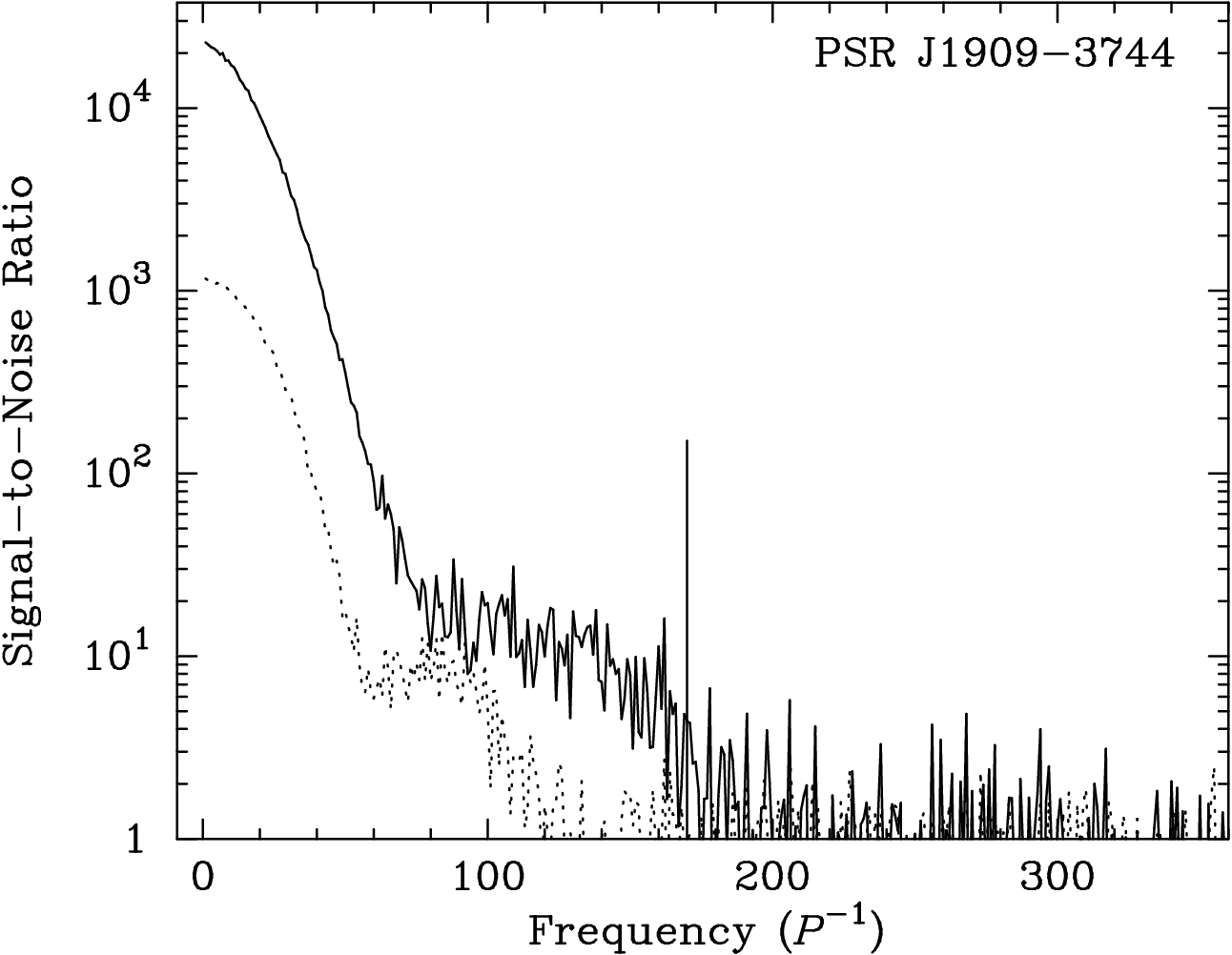} \\
\includegraphics[angle=-90,width=86mm]{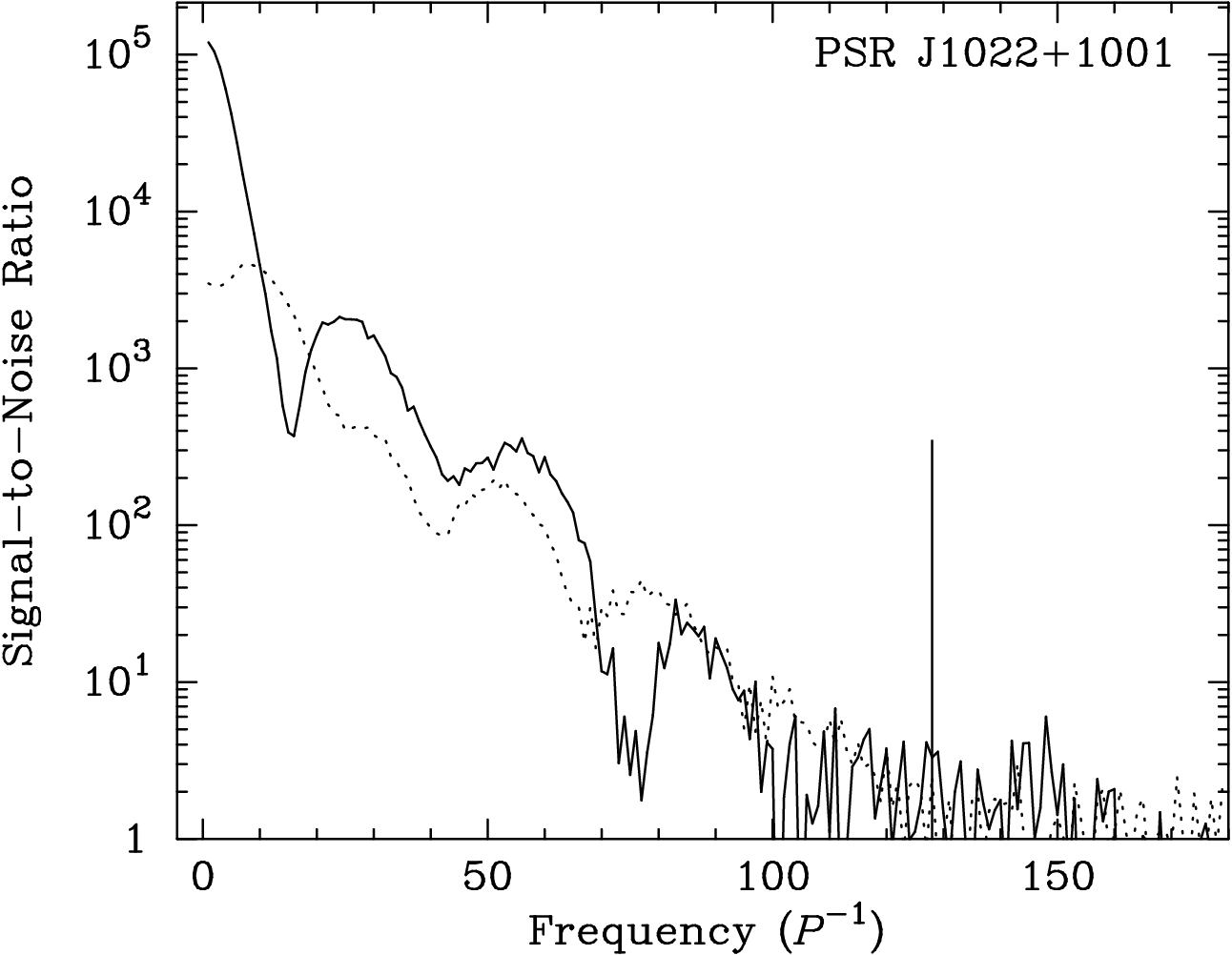} &
\includegraphics[angle=-90,width=86mm]{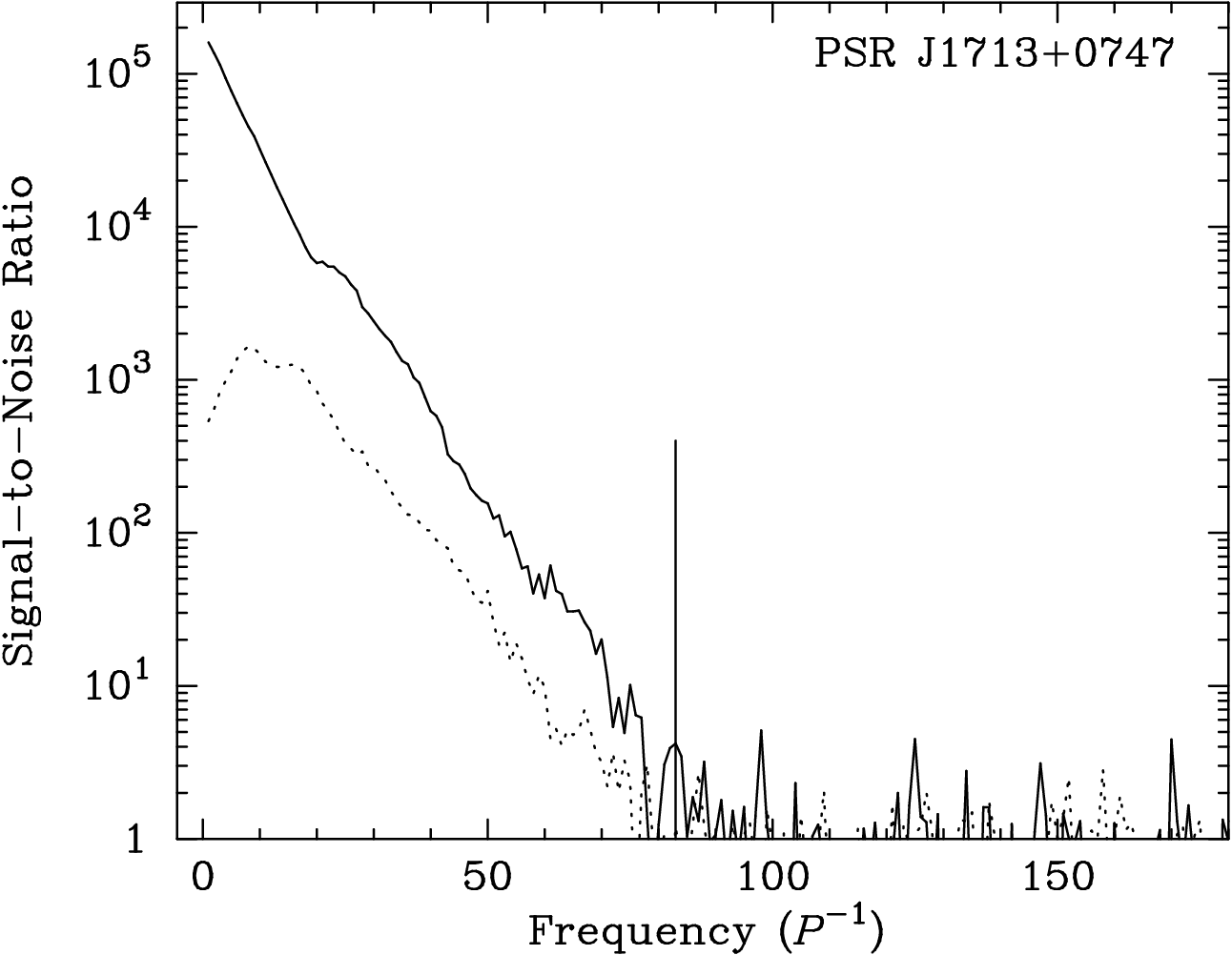} \\
\includegraphics[angle=-90,width=86mm]{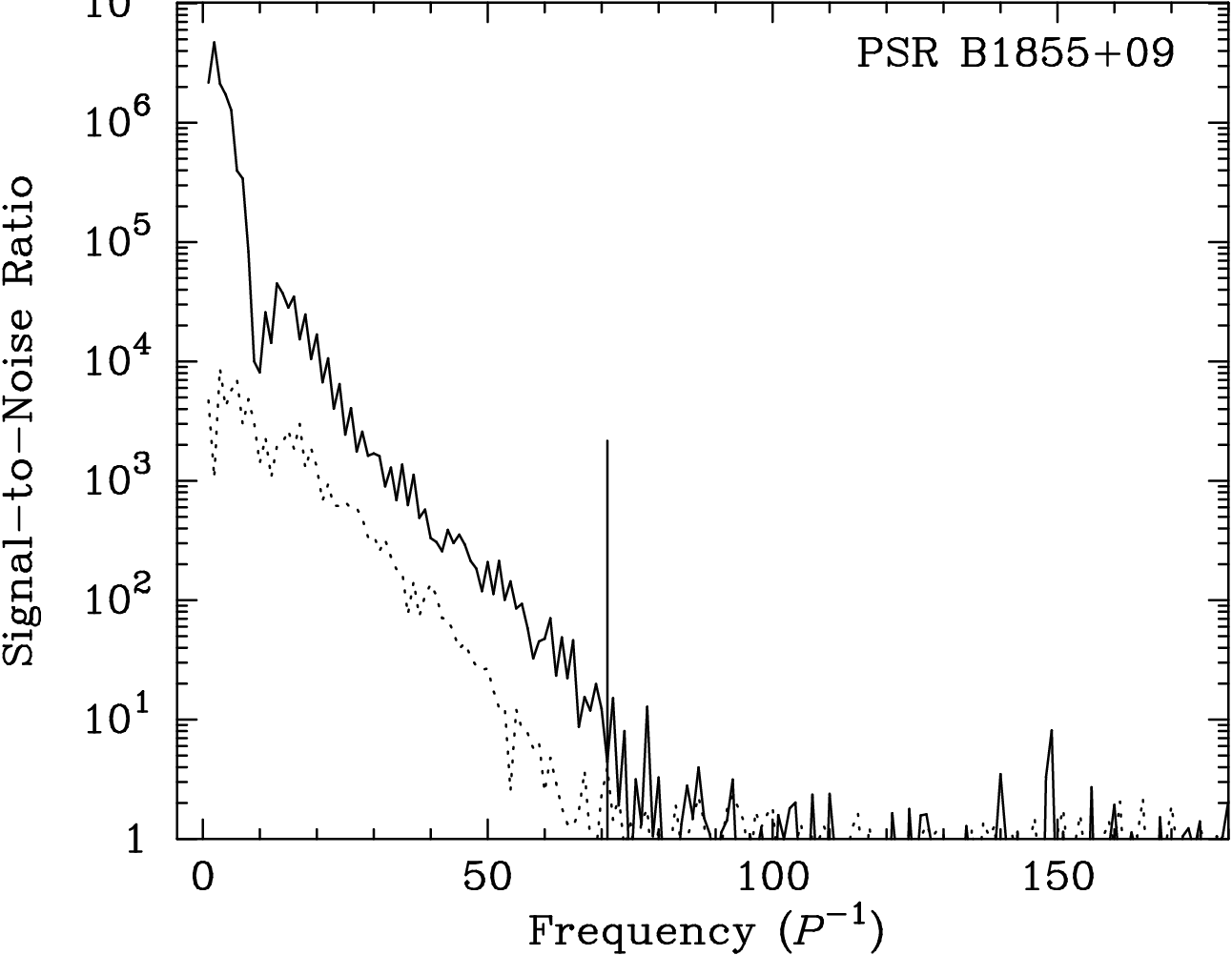} &
\includegraphics[angle=-90,width=86mm]{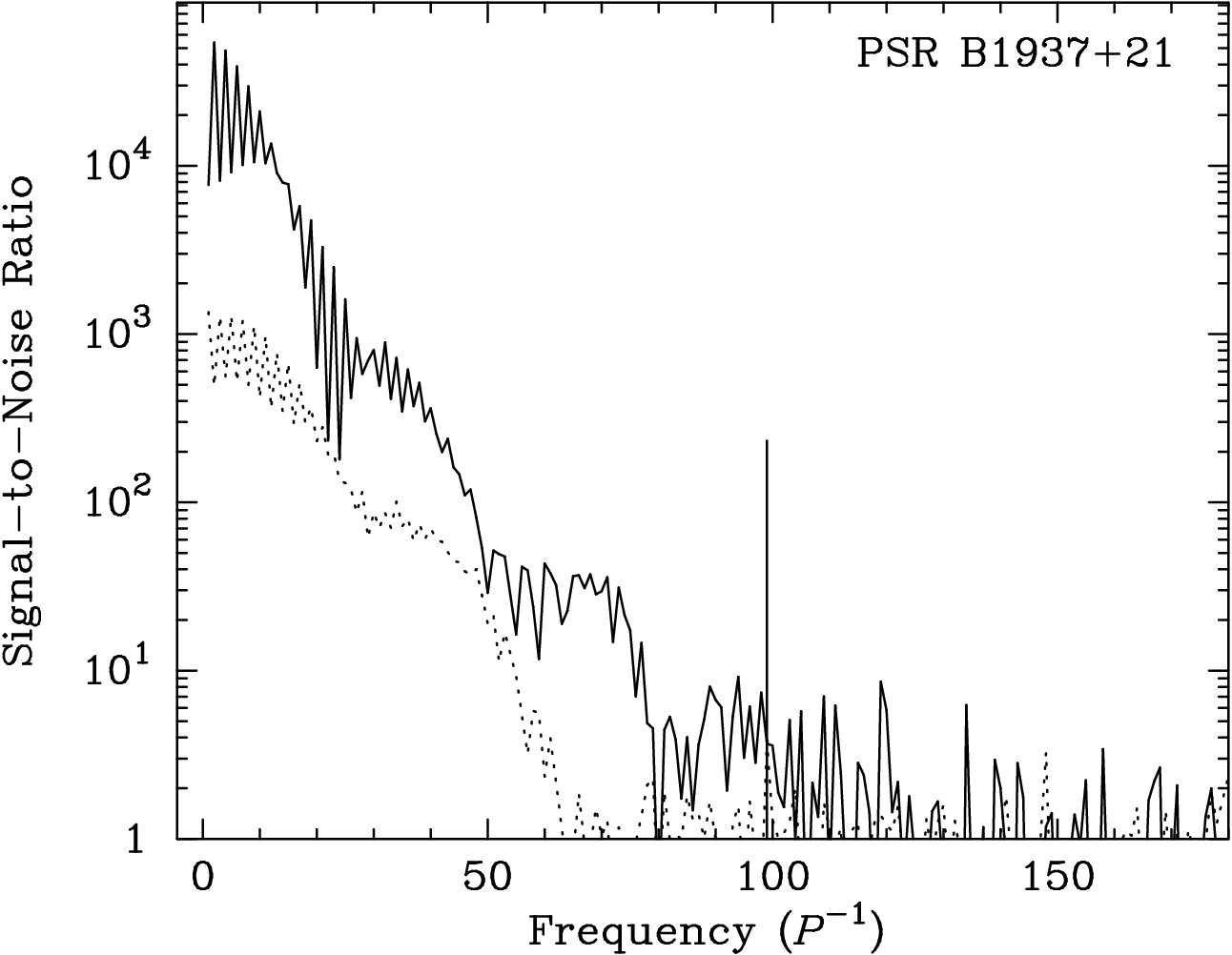}
\end{tabular}
\caption{Fluctuation power spectra of total intensity (solid) and
polarization (dotted).  The vertical line that extends half the height
of each plot indicates the maximum harmonic used to estimate the
arrival time precision. The same number of harmonics are plotted for
each pulsar, except \psr\ and PSR~J1909-3744, for which twice as
many harmonics are required. }
\end{figure*}

The values of $\hat\sigma_{\varphi}$, \Rmult, and \conderr\ (see
\S~\ref{sec:precision}) for each of the selected pulsars are listed in
Table~1.  In all cases, arrival time precision is predicted to improve
through matrix template matching, with timing errors decreasing
between 4\% and 33\%.  As expected, the pulsars with the greatest
improvements (\psr\ and PSR~J1022+1001) are also those with the
greatest amount of polarization fluctuation power (relative to that of
the total intensity) at high frequencies.

% REF 6b
For further comparison, Table~1 also includes
$\hat\sigma_{\tilde\varphi}$, the relative uncertainty of the arrival
time estimate derived by scalar template matching the invariant
profile,
\begin{equation}
S_{\rm inv}(\phi_n) = \left( [S_0(\phi_n)]^2 - |\mbf{S}(\phi_n)|^2 
                      \right)^{1\over2}.
\end{equation}
Again, $\hat\sigma_{\tilde\varphi}$ is normalized with respect to the
uncertainty of the total intensity TOA.  As predicted by a simple
consideration of the noise power in $S_{\rm inv}(\phi_n)$, the
uncertainties of the invariant TOAs are at least $\sqrt{2}$ times
greater than those of the total intensity TOAs.  For PSR~J1022+1001,
the precision yielded by matrix template matching is almost three
times better than that of the invariant TOA.

% REF 2a
The predicted values of $\hat\sigma_{\varphi}$ in Table~1 represent
the minimum increase in experimental sensitivity yielded by matrix
template matching.  They do not include the potentially significant
improvements in timing accuracy that may be gained by using this
technique.  For example, the difference between predicted value for
\psr, $\hat\sigma_{\varphi}=0.8318(1)$, and the measured value of
$\sim 0.74$ reported in~\cite{van05} can be explained by the presence
of systematic timing error in the conventionally-derived TOAs of that
experiment.

\begin{table*}
\caption{Relative Arrival Time Uncertainties}
\begin{center}
\begin{tabular}{lllll}
\tableline
\tableline
Pulsar & \conderr & \Rmult & $\hat\sigma_{\varphi}$ 
% REF 6a
& $\hat\sigma_{\tilde\varphi}$ \\
\tableline
J0437$-$4715 &  0.8285(1) & 0.08993(8) & 0.8318(1) & 1.4314(3) \\
J1022+1001   &  0.615(4)  & 0.394(1)   & 0.669(4)  & 1.80(2)   \\
J1713+0747   &  0.877(2)  & 0.047(3)   & 0.877(2)  & 1.543(5)  \\
B1855+09     &  0.9263(9) & 0.1041(7)  & 0.9314(9) & 1.430(2)  \\
J1909$-$3744 &  0.898(3)  & 0.351(5)   & 0.959(4)  & 1.464(8)  \\
B1937+21     &  0.862(4)  & 0.034(3)   & 0.862(4)  & 1.451(9)  \\
\tableline
\end{tabular}
\tablecomments{Numbers in parenthesis represent the statistical
uncertainty (one standard deviation) in the last digit quoted.  All
errors are normalized by that predicted for conventional pulsar timing
based on the total intensity profile alone. }
\label{tab:precision}
\end{center}
\end{table*}

Timing accuracy is addressed in Table~2, which lists the best
published rms timing residual, $\sigma_\tau$, for each pulsar and the
magnitude of instrumental artifacts that will produce significant
systematic timing error in each data set, as defined by $\Delta\tau
\sim \sigma_\tau$ (see eq.~[\ref{eqn:distortion}]).
% REF 8a
Note that systematic timing errors exist only in the TOAs derived from
the total intensity profile.  The simulation confirms that, within the
experimental uncertainty, there is no distortion of arrival times
derived using matrix template matching, the free transformation {\bf
J} completely absorbs the simulated instrumental boost.

In Table~2, the magnitude of the instrumental error is specified using
the boost impact parameter, $\beta_s$, the equivalent differential
gain ratio, $\gamma_s=2\beta_s$, and the equivalent receptor
non-orthogonality, $\delta_s=2\beta_s$.  Except for PSR~J1713+0747 and
PSR~B1855+09, a conservative value of $\gamma_s\sim3$\% or
$\delta_s\sim2\degr$ results in significant systematic distortion
of arrival times derived by conventional methods.
% REF 2b
Not surprisingly, the two pulsars predicted to be most susceptible to
instrumental error (PSR~J0437$-$4715 and PSR~J1022+1001) are also
those for which the systematic distortion of arrival time estimates
due to poor calibration has already been reported~\citep{van03,hbo04}.

\begin{table*}
\caption{Conventional Systematic Timing Errors}
\begin{center}
\begin{tabular}{lccccc}
\tableline
\tableline
Pulsar & $\sigma_\tau$ (ns) 
       & $\beta_s$ & $\gamma_s$ (\%) & $\delta_s$ (\degr) & ref \\
\tableline
J0437$-$4715 & 130                  & 0.0015 & 0.3  & 0.2 & 5 \\
J1022+1001   & 660\tablenotemark{a} & 0.011  & 2.2  & 1.3 & 1 \\
J1713+0747   & 180                  & 0.13   & 26   & 15  & 4 \\
B1855+09     & 530                  & 0.13   & 26   & 15  & 3 \\
J1909$-$3744 & 74                   & 0.015  & 3.0  & 1.7 & 2 \\
B1937+21     & 140                  & 0.018  & 3.6  & 2.0 & 3 \\
\tableline
\end{tabular}
\tablenotetext{a}{Extrapolated to 60\,min integration.}
\tablerefs{
(1) Hotan, Bailes, \& Ord 2004\nocite{hbo04};
(2) Jacoby et al. 2005\nocite{jhb+05};
(3) Lommen \& Backer 2001\nocite{lb01};
(4) Splaver et al. 2005\nocite{sns+05};
(5) van Straten et al. 2001\nocite{vbb+01}.}
\end{center}
\label{tab:error}
\end{table*}

\section{Instrumental Calibration}
\label{sec:calibration}

In addition to the phase shift, $\varphi$, the matrix template
matching method simultaneously yields the polarimetric response, {\bf
J}, required to transform the observation into the template.  Assuming
that the template has been well-calibrated, this unique transformation
may be applied to completely calibrate the instrumental response in
observations of other point sources. This is demonstrated by matching
an uncalibrated, five minute integration of \psr\ to its standard
template, producing the instrumental parameters shown with their
formal standard deviations in Figure~2.  As in \cite{van04}, the
instrumental response is parameterized by the orientations,
$\theta_k$, and ellipticities, $\epsilon_k$, of the feed receptors,
the differential phase, $\phi$, the differential gain ratio, $\gamma$,
and the absolute gain, $G$, specified in units of an intermediate
reference voltage that can later be calibrated to produce absolute
flux estimates. (Note that in \cite{van04}, $\gamma$ refers to the
boost impact parameter, not the differential gain ratio.)

\begin{figure}
\label{fig:calibration}
% pacv -P -c "0-13" -c "117-127" -D response.eps/cps ../Aspen_2004/pcm.fits
\centerline{\includegraphics[angle=-90,width=86mm]{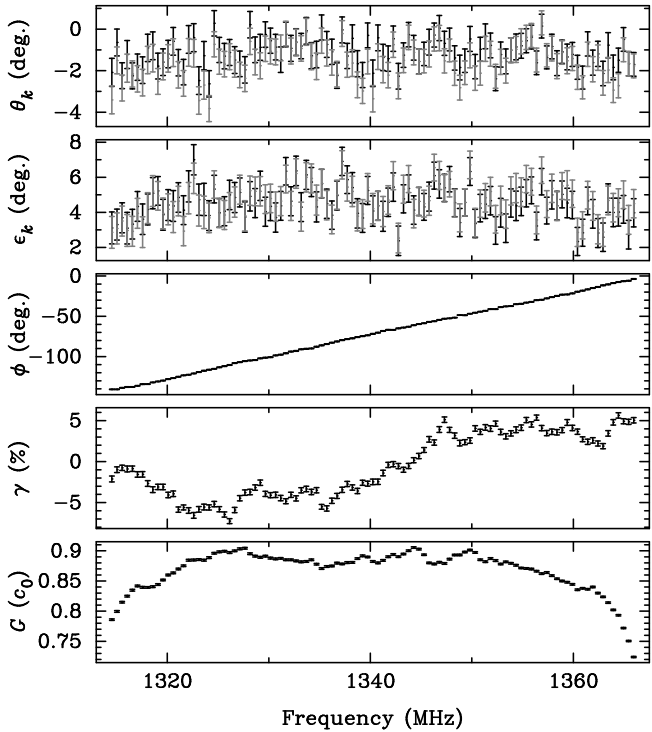}}
\caption{Best-fit instrumental parameters as a function of observing
frequency.  From top to bottom are plotted the orientations,
$\theta_k$, and ellipticities, $\epsilon_k$, of the feed receptors,
the differential phase, $\phi$, the differential gain ratio, $\gamma$,
and the absolute gain, $G$.  In the top two panels, black and gray
correspond to receptors $\mbf{r}_0$ and $\mbf{r}_1$, respectively.}
\end{figure}

In each 500\,kHz channel, it is possible to estimate the ellipticities
and orientations of the feed receptors with an uncertainty of only
$1\degr$.  The apparent correlation between these angles indicates
that the non-orthogonality of the receptors, $\delta_\theta$ and
$\delta_\epsilon$, is better constrained than their absolute
positions.  As a function of frequency, the differential gain ratio,
$\gamma$, exhibits a peak-to-peak variation of about 10\%.  If not
properly calibrated, these deviations will result in severe systematic
distortion of arrival times derived from the total intensity profile
(cf. Table 2).  Furthermore, the effect on the frequency-integrated
profile may vary randomly in time due to interstellar scintillation of
the pulsar signal.  Figure~2 emphasizes the importance of calibrating
with sufficiently high frequency resolution to avoid bandwidth
depolarization \citep{van02}, an irreversible effect that cannot be
modeled in the current formulation of matrix template matching.

\section{Conclusions}
\label{sec:conclusions}

The future of high-precision pulsar timing is inextricably linked with
advances in high-fidelity polarimetry.  For one out of the six
selected pulsars, the improvement in precision yielded by matrix
template matching is better than that produced by doubling the
integration length or instrumental bandwidth.  In the analysis of the
conventional method of scalar template matching, conservative levels
of instrumental distortion are predicted to produce systematic timing
errors of the same order as the rms timing residuals in the current
best data sets.  These errors are completely eliminated by the matrix
template matching method.  
% REF 3d
Therefore, it is expected this technique will perform better than
current conventional methods in the majority of experiments.
Furthermore, the method provides a new means of fully calibrating the
instrumental response using a single observation of a well-determined
pulsar.

Implicit in the application of the matrix template matching method is
the assumption that the average polarization intrinsic to the pulsar
does not vary significantly over any timescale of interest.  Although
a variety of studies have investigated the large fluctuations in the
polarization of single pulses (e.g.\ \cite{edw04}, \cite{kj04}, and
references therein) no research on the long term properties of pulsar
polarization has been published to date.  Most likely, any process
that effects phase-dependent changes in average polarization would
similarly alter the total intensity profile.  Therefore, a systematic
study of the long term stability of millisecond pulsar polarization
would make a valuable contribution toward the major science goals of
this field, such as the detection of low-frequency gravitational
radiation and the verification of relativistic gravity in the
strong-field limit \citep[and references therein]{ckl+04,hob05}.

The impact of polarization on pulsar timing will become increasingly
apparent in a larger number of experiments as instruments with greater
sensitivity are employed, and it is imperative to incorporate a
sophisticated treatment of polarization in high-precision timing
analyses.  All of the software required to perform matrix template
matching is freely available as part of PSRCHIVE~\citep{hvm04}, which
has been openly developed in an effort to facilitate the exchange of
pulsar astronomical data between observatories and research groups.

\acknowledgements

I thank F.~Jenet for contributing valuable advice and feedback
throughout this research, and gratefully acknowledge the thoughtful
criticism of the referee.  Polarization data were provided by
S.~Ord, I.~Stairs, and the data base of published pulse
profiles maintained by the European Pulsar Network.  The Parkes
Observatory is part of the Australia Telescope which is funded by the
Commonwealth of Australia for operation as a National Facility managed
by CSIRO.  This work was supported by NASA under grant NAG5-13396.

%\bibliographystyle{aasjournal}
%\bibliography{journals,modrefs,psrrefs,../local,crossrefs}
%\end{document}

\begin{appendix}

The following additional material was not submitted to The
Astrophysical Journal and was not peer reviewed.  It is provided as
further information for the interested reader. \\ [5mm]

\section{Equation (13): Conditional Variance of $\varphi$}

Equation (13) presents an analytical expression for the conditional
variance of the best-fit phase shift.  Given only the template
profile, it can be used to predict arrival time precision as a
function of the $S/N$ of the observation.  There are two ways to
derive Equation (13):
\begin{enumerate}
\item via Equations (11) and (12); i.e.\ using the $\sin(x)=x$
  small-angle approximation, solving for the slope, and applying
  first-order error propagation; or
\item via Equation (14), which defines the curvature matrix from which
  the uncertainty of the $\varphi$ estimate is formally defined.
\end{enumerate}
Both derivations are shown here.

\subsection{Derivation via Equations (11) and (12)}

Start by deriving Equation (11), noting that Equation (10) can be
written
\begin{equation}
\chi^2 = \sum_{m=1}^{N/2} \sum_{k=0}^3 \varsigma_k^{-2}
\Delta_{k,m} \Delta^*_{k,m},
\label{eqn:merit_delta}
\end{equation}
where
\begin{equation}
\Delta_{k,m}=S_k^\prime(\nu_m)-\trace[\pauli{k}\;\mbf{\rho}^\prime(\nu_m)],
\end{equation}
such that
\begin{equation}
\label{eqn:delchisq_delvarphi_two}
{\partial\chi^2\over\partial\varphi} =
\sum_{m=1}^{N/2} \sum_{k=0}^3 \varsigma_k^{-2}
2 \real \left[ \Delta^*_{k,m} {\partial\Delta_{k,m}\over\partial\varphi} \right].
\end{equation}
Now
\begin{equation}
\label{eqn:delDelta_delvarphi}
{\partial\Delta_{k,m}\over\partial\varphi} = 
i2\pi\nu_m \trace\left[\pauli{k}\;\mbf{\rho}^\prime(\nu_m)\right]
\end{equation}
and $\real[iz]=-\imag[z]$; therefore,
\begin{equation}
\label{eqn:delchisq_delvarphi_Im}
{\partial\chi^2\over\partial\varphi} = -4\pi
\sum_{m=1}^{N/2} \sum_{k=0}^3 \varsigma_k^{-2}
\nu_m \imag \left[ \Delta^*_{k,m} \trace\left[\pauli{k}\;\mbf{\rho}^\prime(\nu_m)\right] \right].
\end{equation}
In the above equation, the second term of $\Delta^*_{k,m}$ is
multiplied by its complex conjugate, resulting in a real number (the
squared modulus) with no imaginary component; therefore,
\begin{equation}
\label{eqn:delchisq_delvarphi_almost}
{\partial\chi^2\over\partial\varphi} = -4\pi
\sum_{m=1}^{N/2} \sum_{k=0}^3 \varsigma_k^{-2}
\nu_m \imag \left[ S_k^{\prime*}(\nu_m) \trace\left[\pauli{k}\;\mbf{\rho}^\prime(\nu_m)\right] \right].
\end{equation}
Now define the cross-spectral power of the template and the observation
\begin{equation}
\label{eqn:cross_spectral_power}
S_{k,m}=S_k^{\prime*}(\nu_m)\trace\left[\pauli{k}\;\mbf{\rho}_0(\nu_m)\right],
\end{equation}
such that
\begin{equation}
\label{eqn:delchisq_delvarphi_cross}
{\partial\chi^2\over\partial\varphi} = -4\pi
\sum_{m=1}^{N/2} \sum_{k=0}^3 \varsigma_k^{-2}
\nu_m \imag \left[ S_{k,m} \exp(-i2\pi\nu_m\varphi) \right].
\end{equation}
Equation (11) is obtained after expressing $S_{k,m}$ in polar
coordinates; Equation (12) is a small angle (or high $S/N$)
approximation.  Note that the opening minus sign is missing in these
equations. \\ [5mm]

\noindent
To derive Equation (13), first solve Equation (12) for $\varphi$
\begin{equation}
\label{eqn:varphi}
\varphi = 
{ \mbf{\sum} \varsigma_k^{-2}|S_{k,m}|\phi_{k,m}\nu_m \over
  2\pi \mbf{\sum} \varsigma_k^{-2}|S_{k,m}|\nu_m^2 },
\end{equation}
where $\mbf{\sum}=\sum_{m=1}^{N/2} \sum_{k=0}^3$ is introduced for
convenience.  Standard error propagation is greatly simplified by
recognizing that the variance of $\phi_{k,m}$ is $|S_{k,m}|$.  Therefore,
to first order, the variance of $\varphi$ is given by
\begin{equation}
\var(\varphi)= \sum_{m=1}^{N/2}\sum_{k=0}^3 
  \left({\partial\varphi\over\partial\phi_{k,m}}\right)^2 \varsigma_k^{-2}|S_{k,m}|
\label{eqn:varvarphi}
\end{equation}
Substitution of
\begin{equation}
{\partial\varphi\over\partial\phi_{k,m}} 
  = { \varsigma_k^{-2}|S_{k,m}|\nu_m \over
  2\pi \mbf{\sum} \varsigma_k^{-2}|S_{k,m}|\nu_m^2 }
\label{eqn:partialvarphi}
\end{equation}
into equation~\ref{eqn:varvarphi} yields
\begin{equation}
\var(\varphi)= \sum_{m=1}^{N/2}\sum_{k=0}^3 { \varsigma_k^{-2}|S_{k,m}| \nu_m^2 
\over 
4\pi^2 \left( \mbf{\sum} \varsigma_k^{-2}|S_{k,m}|\nu_m^2  \right)^2 } 
= {1\over4\pi^2 \mbf{\sum} \varsigma_k^{-2}|S_{k,m}|\nu_m^2},
\label{eqn:intermediate}
\end{equation}
which, after expanding $\mbf{\sum}$, yields Equation (13).

\subsection{Derivation via Equation (14)}

When no other parameters are varied, the conditional variance,
\begin{equation}
  \var(\varphi|{\bf J}) = c_{\varphi\varphi} = \alpha_{00}^{-1}
\end{equation}
where $\alpha$ is the $1\times1$ curvature matrix defined by Equation (14) and
\begin{equation}
\alpha_{00} = {1\over2}{\partial^2\chi^2\over\partial\varphi\partial \varphi} 
= {2 \over \varsigma^2} \sum_{m=1}^{N/2}
  \real\left[
    \trace\left( {\partial\mbf{\rho}^{\prime\dagger}_m\over\partial\varphi}
                 {\partial\mbf{\rho}^\prime_m\over\partial\varphi} 
          \right)
          \right].
  \label{eqn:alpha_varphi}
\end{equation}
Using
\begin{equation}
  {\partial\mbf{\rho}^\prime_m\over\partial\varphi}
  = -i2\pi\nu_m \mbf{\rho}^\prime_m
\end{equation}
and
\begin{equation}
  \trace(\mbf{\rho}^\dagger\mbf{\rho}) = {1\over2}\sum_{k=0}^3 S_k S_k^*,
\end{equation}
\Eqn{alpha_varphi} becomes
\begin{equation}
\alpha_{00} = {4\pi \over \varsigma^2} \sum_{m=1}^{N/2} \nu_m^2 \sum_{k=0}^3 |S_{k,m}|
\end{equation}
which, in turn, yields Equation (13) if it is assumed that $\varsigma_k=\varsigma$.

\subsection{Interpretation}

\begin{enumerate}

\item Equation (13) is the Fourier domain representation of Equation (B1) of \citet{dr83}; this can be proven by noting that, if $X(\nu)$ is the Fourier transform of $x(\phi)$, then $i2\pi \nu X(\nu)$ is the Fourier transform of $dx/d\phi$ and, by Parseval's Theorem, the integral of $[dx/d\phi]^2$ in the phase domain is equal to the integral of $\nu^2|X(\nu)|^2$ in the frequency domain.

\item Phases such as $\phi_n=n/N$ and $\varphi$ are dimensionless turns;
therefore, TOA $\tau=\tau_0+\varphi P$, where $P$ is the pulse period.

\item Frequencies such as $\nu_m=m$ are dimensionless harmonics.

\item $S_{k,m}$ is the cross spectral power between the
template and profile; in the case of the theoretical prediction, it is
the autospectral power in the template.

\item The r.m.s.\ of the noise in the Fourier domain $\varsigma_k$
is a function of the $S/N$ of the observation.

\item Equation (13) returns variance in dimensionless
phase, which can be translated into arrival time error with
$\sigma_\tau = \sqrt{\var(\varphi)} P$.

\item Based on the approximation, $\sigma_\tau = w (S/N)^{-1}$, were $w$ is
  the width of the pulse, Equation (13) can be used to define the effective
  width of a pulse.
  
\end{enumerate}

\section{Equation (14): Curvature Matrix}

Begin with the definition of the merit function,
\begin{equation}
\chi^2 = \sum_{m=1}^{N/2} \sum_{k=0}^3 { |S_k^\prime(\nu_m) -
\trace[\pauli{k}\,\mbf{\rho}^\prime(\nu_m)]|^2 \varsigma_k^{-2} }.
\label{eqn:merit_two}
\end{equation}
Letting $\mbf{\rho}^\prime_m = \mbf{\rho}^\prime(\nu_m)$ and 
$\varsigma_k=\varsigma$, the first
partial derivative is
\begin{equation}
{\partial\chi^2\over\partial\eta_r} = -{2\over\varsigma^2} 
\sum_{m=1}^{N/2} \sum_{k=0}^3 
\real\left[ (S_k^\prime(\nu_m) - \trace[\pauli{k}\,\mbf{\rho}^\prime_m])
        \trace\left(\pauli{k}\,{\partial\mbf{\rho}^\prime_m\over\partial\eta_r}
             \right)^*
  \right] 
\label{eqn:first}
\end{equation}
Because the Pauli spin matrices are Hermitian,
$\trace(\pauli{k}\,\mbf{\rho})^*=\trace(\pauli{k}\,\mbf{\rho}^\dagger)$.
The second partial derivative is
\begin{equation}
{\partial^2\chi^2\over\partial\eta_r\partial\eta_s} = {2\over\varsigma^2} 
\sum_{m=1}^{N/2} \sum_{k=0}^3 
\real\left[ 
 \trace\left(\pauli{k}\,{\partial\mbf{\rho}^\prime_m\over\partial\eta_s}\right)
 \trace\left(\pauli{k}\,{\partial\mbf{\rho}^{\prime\dagger}_m\over\partial\eta_r}\right)
  \right].
\label{eqn:second}
\end{equation}
Following the discussion in \S\,15.5 of Numerical Recipes, the
term containing a second derivative in \eqn{curvature} has been
dropped.  Now the trace of a matrix is a scalar and
$c\,\trace({\bf A})=\trace(c{\bf A})$; therefore,
\begin{equation}
{\partial^2\chi^2\over\partial\eta_r\partial\eta_s} = {2\over\varsigma^2} 
\sum_{m=1}^{N/2} \sum_{k=0}^3 
\real\left[\trace\left(
 \trace\left[\pauli{k}\,{\partial\mbf{\rho}^\prime_m\over\partial\eta_s}\right]
 \pauli{k}\,{\partial\mbf{\rho}^{\prime\dagger}_m\over\partial\eta_r}\right)
  \right] 
\label{eqn:linear1}
\end{equation}
Furthermore $\trace({\bf A})+\trace({\bf B}) = \trace({\bf A}+{\bf B})$
and $\real(z)+\real(w)=\real(z+w)$, so
\begin{equation}
{\partial^2\chi^2\over\partial\eta_r\partial\eta_s} = {2\over\varsigma^2} 
\sum_{m=1}^{N/2} 
\real\left[\trace\left(\left[\sum_{k=0}^3
 \trace\left(\pauli{k}\,{\partial\mbf{\rho}^\prime_m\over\partial\eta_s}\right)
 \pauli{k}\right]
 {\partial\mbf{\rho}^{\prime\dagger}_m\over\partial\eta_r}\right)
  \right].
\label{eqn:linear2}
\end{equation}
The first factor inside the trace,
\begin{equation}
\sum_{k=0}^3
 \trace\left(\pauli{k}\,{\partial\mbf{\rho}^\prime_m\over\partial\eta_s}\right)
 \pauli{k} =  2{\partial\mbf{\rho}^\prime_m\over\partial\eta_s}
\end{equation}
and $\trace({\bf AB}) = \trace({\bf BA})$; therefore
\begin{equation}
\label{eqn:curvature_two}
{\partial^2\chi^2\over\partial\eta_r\partial \eta_s} 
= {4 \over \varsigma^2} \sum_{m=1}^{N/2}
  \real\left[
    \trace\left( {\partial\mbf{\rho}^{\prime\dagger}_m\over\partial\eta_r}
                 {\partial\mbf{\rho}^\prime_m\over\partial\eta_s} 
          \right)
       \right].
\end{equation}

\section{Equation (28): Non-orthogonality, $\sin(\delta)$}

Begin with
\begin{equation}
\label{eqn:cos_delta}
\cos2\delta = -\cos\Theta 
= - {\mbf{S}_0\mbf{\cdot}\mbf{S}_1 \over |\mbf{S}_0||\mbf{S}_1|}
= 1-2\left({|\mbf{r}_0^\dagger\mbf{r}_1|\over|\mbf{r}_0||\mbf{r}_1|}\right)^2
\end{equation}
then use the double-angle formula $\cos2\delta = 1-2\sin^2\delta$.

\end{appendix}
 
\end{document}